\documentclass{statsoc}
\usepackage[a4paper, left=3cm,right=3cm,top=3cm,bottom=3cm]{geometry}
\usepackage{amsmath}
\graphicspath{{figures/}}
\usepackage{float}
\usepackage{dsfont}
\usepackage{amsfonts}
\usepackage[T1]{fontenc}
\usepackage[ruled,linesnumbered]{algorithm2e}
\usepackage{xcolor}
\usepackage{stmaryrd}

\usepackage{comment}

\usepackage{url}
\newcommand{\e}[1]{\mathbb{E}\left[#1\right]  }
\newcommand{\indep}{\perp\!\!\!\perp}
\usepackage{subfig}
\usepackage{graphicx}

\usepackage{marvosym}
\usepackage{xspace}

\usepackage{thm-restate}

\usepackage{graphicx}

\usepackage{shadethm}
\usepackage{amssymb}
\usepackage{amsmath}

\usepackage{todonotes}

\newtheorem{theorem}{Theorem}

\newtheorem{definition}{Definition}
\newtheorem{proposition}{Proposition}

\newtheorem{lemma}[theorem]{Lemma}

\newcommand{\Comment}[1]{\textcolor{blue}{\textbf{\texttt{// #1}}}}

\title{When knockoffs fail: diagnosing and fixing non-exchangeability of knockoffs}

\author[Alexandre Blain {\it et al.}]{Alexandre Blain}\address{Université Paris-Saclay, Inria, CEA, Palaiseau, 91120, France.}

\author{Angel Reyero Lobo}\address{Institut de Mathématiques de Toulouse ; UMR5219; Université de Toulouse ; CNRS ; UPS, F-31062 Toulouse Cedex 9; Université Paris-Saclay, Inria, CEA, Palaiseau, 91120, France.}\email{angel.reyero-lobo@inria.fr}

\author{Julia Linhart}\address{Université Paris-Saclay, Inria, CEA, Palaiseau, 91120, France.}

\author{Bertrand Thirion}\address{Université Paris-Saclay, Inria, CEA, Palaiseau, 91120, France.}

\author[Blain {\it et al.}]{Pierre Neuvial}\address{Institut de Mathématiques de Toulouse ; UMR5219; Université de Toulouse ; CNRS ; UPS, F-31062 Toulouse Cedex 9, France.}

\usepackage{hyperref}
\hypersetup{
    colorlinks=true,
    linkcolor=purple,   
    citecolor=purple,
    filecolor=purple,
    urlcolor=purple,
    linktocpage=true
}

\begin{document}

\begin{abstract}
Knockoffs are a popular statistical framework that addresses the challenging problem of conditional variable selection in high-dimensional settings with statistical control. 
Such statistical control is essential for the reliability of inference.
However, knockoff guarantees rely on an exchangeability assumption that is difficult to test in practice, and there is little discussion in the literature on how to deal with unfulfilled hypotheses. 
This assumption is related to the ability to generate data similar to the observed data. 
To maintain reliable inference, we introduce a diagnostic tool based on Classifier Two-Sample Tests.
Using simulations and real data, we show that violations of this assumption occur in common settings for classical knockoff generators, especially when the data have a strong dependence structure. 
We show that the diagnostic tool correctly detects such behavior.
We show that an alternative knockoff construction, based on constructing a predictor of each variable based on all others, solves the issue. 
We also propose a computationally-efficient variant of this algorithm and show empirically that this approach restores error control on simulated data and semi-simulated experiments based on neuroimaging data.
\end{abstract}


\section{Introduction}
Controlled variable selection is a fundamental problem encountered in diverse fields where practitioners have to assess the importance of input variables with regards to an outcome of interest. 
Variable selection aims at choosing variables that should be included in an analysis or model while excluding those that do not substantially contribute to the understanding or prediction of an outcome.
Statistically controlled variable selection aims at limiting the proportion of selected variables that are independent of the outcome. This problem is much harder than usual variable selection since guarantees on False Positives have to be provided.
For instance, selecting variables is useful in epidemiology to better understand driving factors in the propagation of diseases \citep{greenland1989modeling}. In mathematical finance, choosing an optimal portfolio amounts to solving a variable selection problem \citep{ait2001variable}. In neuroimaging, practitioners are interested in associating cognitive functions to relevant brain regions \citep{weichwald2015causal}. In genomics, identifying gene variants that drive disease outcomes is crucial \citep{visscher2012five}. Additionally, False Discoveries have to be controlled to ensure that variable selection is reliable \citep{benjamini1995controlling, genovese2004stochastic}.\\

\begin{figure}
\centering
\includegraphics[width=0.95\linewidth]{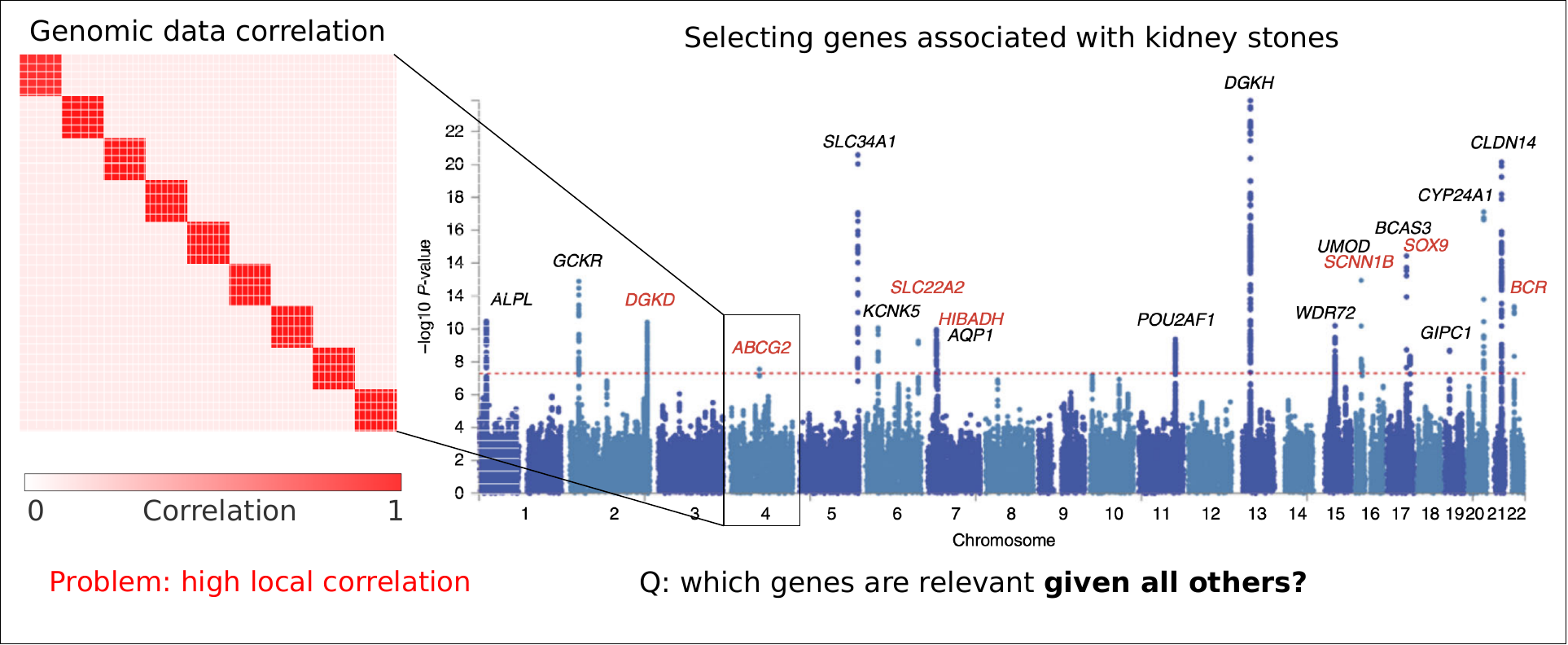}
\caption{ \textbf{Illustration of conditional variable selection.} Given a genomic dataset and kidney stones diagnoses (outcome), we aim at finding variables that add information w.r.t the outcome given all other variables. This problem is challenging since genes are highly correlated locally. Assessing the added information of specific genes given all others is crucial to improve the understanding of disease outcomes. Details about this Manhattan plot are available in \citealp{howles2019genetic}.}
\label{fig:conditional}
\end{figure}

In conditional variable selection, the objective is to assess the value of a particular variable w.r.t. an outcome given a particular set of other variables \citep{konig2021relative}.
This approach differs markedly from classical marginal inference, where variables are considered individually in relation to the outcome \citep{candes2018panning}. 
With conditional variable selection, we examine whether a variable remains informative and relevant when considered alongside other variables. 
This is particularly crucial when dealing with datasets characterized by high levels of correlation among variables. 
Inference in the presence of correlations is illustrated with a genomics example of Genome-Wide Association Study (GWAS) in Figure \ref{fig:conditional}, where correlation between variables is due to a biological phenomenon called linkage disequilibrium \citep{uffelmann2021genome-wide}.\\

This problem becomes difficult when there is a high level of correlation between variables on which one aims to perform inference. 
In machine learning contexts, the difficulty of identifying features that add unique predictive value when combined with others in large-correlation contexts has been well identified \citep{bickel2009simultaneous, mandozzi2016hierarchical, goeman2007analyzing, javanmard2014confidence}.
High correlations are ubiquitous in many application areas, for example when the features of interest are derived from some biological measurements that inherently have complex dependencies.
For instance, in brain mapping, practitioners may want to detect brain regions that are relevant for a given cognitive task given the rest of the brain \citep{weichwald2015causal}, but have to deal with the strong dependencies observed in these data \citep{chevalier2021decoding}.
In genomics, understanding which genes conditionally affect disease outcomes can lead to more effective disease prevention and intervention strategies \citep{sesia2019gene}.\\

Knockoffs \citep{barber2015controlling, candes2018panning} have emerged as a popular and powerful tool for addressing the problem of conditional variable selection with guarantees on the rate of false detections. 
This framework consists of guiding variable selection by generating synthetic variables, referred to as \emph{knockoffs}, that closely mimic the statistical properties of the original variables while not being associated with the outcome of interest (conditionally on their original counterpart). 
This approach aims to identify genuine associations between variables and outcomes by comparing them with their corresponding knockoffs.
An attractive feature of knockoffs is that inference is performed simultaneously for all variables, resulting in an acceptable computational cost. Performing the same type of inference for each variable leads to as many Conditional Randomization Tests \citep{candes2018panning, Nguyen2022CRT} as the number of covariates, which has a high computational cost.\\

It is important to note, however, that the error control of knockoffs relies on a critical assumption called \textit{exchangeability}. 
For exchangeability to hold, the joint distribution of the data must remain unchanged when an original variable is exchanged for its knockoff counterpart.
While knockoffs have shown promise in many applications, this assumption of exchangeability requires careful consideration and assessment, as its validity impacts the reliability of the entire variable selection process. 
While the generative processes could ideally be known or derived from first principles, for many practical problems, they are unknown or intractable ; current practice then  relies on learning the joint distributions of the observations in order to draw knockoffs from an appropriately perturbed model.\\

The most popular knockoff generation procedure involves a Gaussian assumption, and thus relies on the knowledge or the accurate estimation of the covariance structure of the covariates, which leads to two potential issues: \textit{i)} the violation of the Gaussian hypothesis and \textit{ii)} inaccuracies in the covariance estimation. This worsens with the number of variables $p$ as the problem becomes harder, especially in high-dimensional regimes where the number of samples $n$ is comparatively small.

Some alternative knockoff generation methods avoid the problem of covariance estimation via e.g. Deep Learning \citep{romano2020deep}. However, such methods also suffer from high-dimensional regimes as deep neural networks require massive amounts of data to be properly trained, especially in large feature spaces. Overall, methods that require a large sample size $n$ and a small number of variables $p$ may not be adapted to the knockoffs framework, which is designed for high-dimensional variable selection.
Domain-specific procedures have been developed to tackle some of these issues. 
In particular, motivated by applications to GWAS in genomics, Monte-Carlo based knockoff generation methods have been proposed for discrete distributions \citep{sesia2019gene, bates2021metropolized}.
In the continuous setting, the latter is equivalent to Gaussian knockoffs.\\

This paper is organized as follows.
In Sections \ref{sec:knockoff-framework} and~\ref{sec:genKO}, we describe the knockoff framework and review knockoff generation methods.
In particular, an efficient nonparametric algorithm for constructing knockoff variables is proposed in Section \ref{sec:NGKO}. 
In Section \ref{sec:KOfail}, we introduce an effective diagnostic tool that allows practitioners to examine knockoffs along the original data for potential violations. 
In Section \ref{sec:results} this tool is implemented to illustrate through simulations and application to semi-simulated fMRI data common cases of non-exchangeability when using Gaussian knockoffs and associated error control violations.
We also show that the proposed nonparametric knockoff generation procedure restores error control in all the cases considered.
%
%

\section{Knockoff framework}
\label{sec:knockoff-framework}

\textbf{Notation.}
For an integer $k$, $\llbracket k \rrbracket$ denotes the set $\{1, \dots, k\}$. Equality in distribution is denoted by $\stackrel{d}{=}$.
%
%
A vector $X = (X_j)_{j \in \llbracket p \rrbracket}$ from which the $j^{th}$ coordinate is removed is denoted by $X_{-j}$, that is, $ X 
\setminus \{X_j\}$. The independence between two random vectors
$X$ and $Y$ is denoted by $X \perp
Y$. 
For two vectors $X$ and $
\tilde{X}$ of size $p$ and a subset $S \subset \llbracket p \rrbracket$,  $(X,
\tilde{X})_{swap(S)}$ denotes the vector obtained from
$(X, \tilde{X})$ by swapping the entries $X_j$ and
$\tilde{X}_j$ for each $j \in S$. 
Matrices are denoted by bold letters.
$\mathbf{X}_j$ (resp. $\mathbf{X}^i$) stands for the $j$-th column (resp. $i$-th row) of matrix $\mathbf{X}$.
For any set $S$, $|S|$ denotes the cardinality of $S$. 
Regressors will be taken from $\mathcal{F}$, which denotes a class of measurable functions. 
\textbf{Problem setup.} The input data are denoted by $ \mathbf{X} \in \mathbb{R}^{n \times p}$, where $n$ is the number of samples and $p$ the number of variables. The outcome of interest is denoted by $\mathbf{Y} \in \mathbb{R}^{n}$. 
The goal is to select variables that are relevant with respect to the outcome \emph{conditionally on all others}. Formally, we test simultaneously for all $j \in \llbracket p\rrbracket$:
\begin{align}
H_{0, j}: Y \perp X_j | X_{-j} 
\quad \text{versus} \quad 
H_{1, j}: Y \not\perp X_j | X_{-j}.\nonumber
\end{align}
The output of a variable selection method is a rejection set $\hat{S}
\subset \llbracket p\rrbracket$ that estimates the true unknown support
$\mathcal{H}_{1} = \{j : Y \not\perp X_j \mid X_{-j} \}$. Its
complement is the set of true
null hypotheses $\mathcal{H}_0 = \{j : Y \perp X_j \mid X_{-j}
\}$.
To ensure reliable inference, our aim is to provide a statistical guarantee on the proportion of False Discoveries in $\hat{S}$. The False Discovery Proportion (FDP) and the False Discovery Rate (FDR) \cite{benjamini1995controlling} are defined as:
\begin{align}
\mathrm{FDP}(\hat{S})=\frac{|\hat{S} \cap \mathcal{H}_{0}|}{|\hat{S}| \vee 1}, \quad \mathrm{FDR}(\hat{S})=\mathbb{E}[\mathrm{FDP}(\hat{S})].\nonumber%
\end{align}\\

\textbf{Knockoffs.} Knockoff inference leverages noisy duplicates known as knockoff variables, which serve as controls in the variable selection process. A key challenge in this method is to ensure that these knockoff variables maintain the same correlation structure as the original variables while being conditionally independent of the outcome. This is essential to allow meaningful comparisons between the original variables and their knockoff counterparts, thereby identifying variables that provide relevant information regarding the outcome. Knockoff variables are defined as follows:

\begin{restatable}[Model-X knockoffs, \citealp{candes2018panning}]{definition}{kodef}
\label{def:MXKnockoffs}
For the family of random variables $X = (X_1, \ldots, X_p)$, knockoffs are a new family of random variables $\tilde{X} = (\tilde{X}_1, \ldots,\tilde{X}_p)$ satisfying:
\begin{enumerate}
\item for any $S \subset \llbracket p \rrbracket$, $(X,  \tilde{X})_{swap(S)} \stackrel{d}{=} (X, \tilde{X})$\\
\item $\tilde{X} \perp Y| X$.
\end{enumerate}
\end{restatable}
\textbf{Exchangeability} is the first property of the above definition. For any set of swapped variables, the joint distribution of real and knockoff variables must remain identical to the original one. Building knockoff variables that satisfy this property is a difficult problem in itself -- we discuss existing work on knockoff generation in Section \ref{sec:genKO}. The second property is generally achieved by building $\tilde{X}$ from $X$ only.\\

\textbf{Statistical inference on knockoff variables.} Once valid knockoffs have been created, conditional variable selection can be performed. To distinguish variables that are substantially more important than their corresponding knockoffs, a machine learning model is employed to generate importance scores for each variable and its respective knockoff. This step enables the identification of variables that offer valuable insights into the outcome, as they exhibit significant disparities in importance compared to their knockoffs. Quantitatively, this is done by computing knockoff statistics $W = (W_1, \ldots, W_p)$ that are defined as follows.\\

\begin{definition}[Knockoff statistic, \citealp{candes2018panning}]
  \label{def:KO-stat}
  A knockoff statistic $W = (W_1, \ldots, W_p)$ is a measure of feature importance that satisfies:
  \begin{enumerate}
  \item $W$ depends only on $\mathbf{X}, \mathbf{\tilde{X}}$ and $\mathbf{Y}$:
  $W = g(\mathbf{X}, \mathbf{\tilde{X}}, \mathbf{Y})$.\\
  \item Swapping column $\mathbf{X}_j$ and its knockoff column $\mathbf{\tilde{X}}_j$ switches the sign of $W_j$: 
    \begin{equation*}
      W_j([\mathbf{X}, \mathbf{\tilde{X}}]_{swap(S)}, \mathbf{Y}) =
      \left\{
          \begin{array}{ll}
            W_j([\mathbf{X}, \mathbf{\tilde{X}}], \mathbf{Y}) \ \text{if } j \in S^{c} \\
            -W_j([\mathbf{X}, \mathbf{\tilde{X}}], \mathbf{Y}) \ \text{if } j \in S . \\
          \end{array}
      \right.
    \end{equation*}
  \end{enumerate}
\end{definition}

The most commonly used knockoff statistic is the Lasso-coefficient difference (LCD) \citep{weinstein2020power}. This statistic is obtained by fitting a Lasso estimator \citep{tibshirani1996regression} on $[\mathbf{X}, \mathbf{\tilde{X}}] \in \mathbb{R}^{n \times 2p}$, which yields $\widehat{\beta} \in \mathbb{R}^{2p}$. Then, the knockoff statistic can be computed using $\widehat{\beta}$:
\begin{align}
\forall j  \in \llbracket p \rrbracket, \quad W_j = |\widehat{\beta}_j| - |\widehat{\beta}_{j+p}|.\nonumber
\end{align}
This coefficient summarizes the importance of the original $j^{th}$ variable relative to its own knockoff: $W_j > 0$ indicates that the original variable is more important for fitting $Y$ than the knockoff variable, meaning that the $j^{th}$ variable is likely relevant. Conversely, $W_j < 0$ indicates that the $j^{th}$ variable is probably irrelevant.
We thus wish to select variables corresponding to large and positive $W_j$. Formally, the rejection set $\hat{S}$ of the knockoffs method can be written $\hat{S} = \{j : W_j > T_q\},$ where:
$$
T_q = \min \left\{t>0: \frac{1+\#\left\{j: W_j \leqslant-t\right\}}{\#\left\{j: W_j \geqslant t\right\}} \leqslant q\right\}.
$$

This definition of $T_q$ ensures that the $\operatorname{FDR}$ is controlled at level $q$ \citep{candes2018panning}. Alternatively, inference can be performed using $\pi$-statistics, which quantify the evidence against each variable:

\begin{equation}
\pi_{j}=\left\{\begin{array}{l}
\frac{1+|\left\{k: W_{k} \leq-W_{j}\right\}|}{p} \text { if } \quad W_{j} > 0\\
1 \quad \text { if } \quad W_{j} \leq 0.
\end{array}\right.
\label{eq:proportion}
\end{equation}

\begin{figure}
\centering
\includegraphics[width=\linewidth]{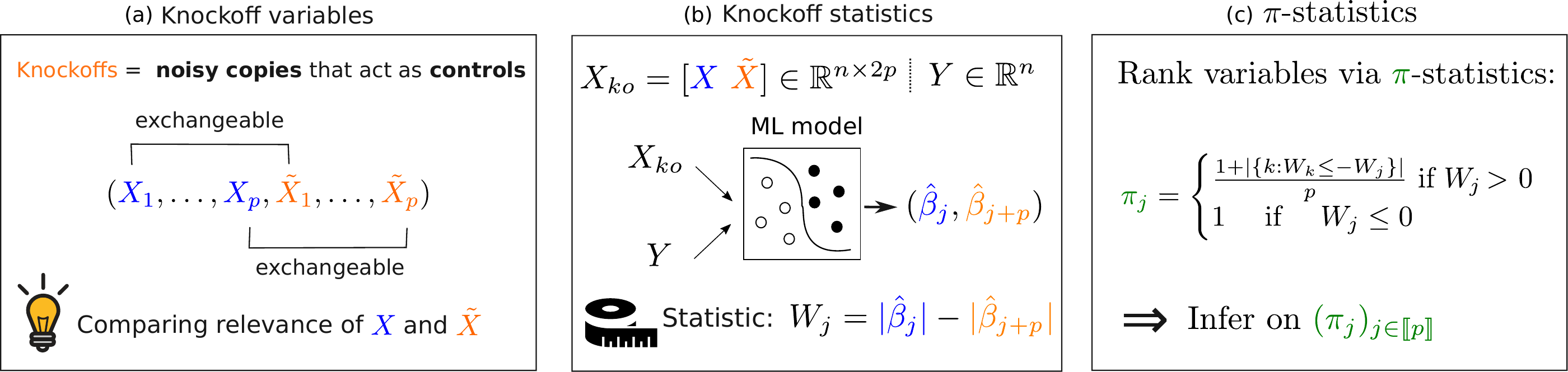}
\caption{\textbf{Conditional variable selection via knockoff variables.} Knockoff variables are noisy copies of the original variables that are used as controls for variable selection (Panel a). Conditionally on the real variables, knockoffs are independent of the outcome $y$. Importance scores for each variable are computed using the sensitivity estimate from machine learning  model -- typically, the corresponding coefficient of a Lasso fit. The knockoff statistic is defined as the difference of modulus between the score associated with the real variable and its knockoff counterpart (Panel b). Then, $\pi$-statistics are computed to rank variables and to perform inference (Panel c).}
\label{fig:koprinciple}
\end{figure}

As noted by \cite{nguyen2020aggregation}, the vanilla knockoffs procedure of \citealp{candes2018panning} is equivalent to using the \cite{benjamini1995controlling} procedure at level $q$ on the vector of $\pi$-statistics $(\pi_j)_{j \in \llbracket p \rrbracket}$. The complete procedure is summarized in Figure \ref{fig:koprinciple}.\\

\section{Knockoff generation methods}
\label{sec:genKO}
\subsection{Gaussian knockoffs}
\citealp{candes2018panning} have introduced a knockoff construction framework for Gaussian data. 
Given $X \sim \mathcal{N}(0, \boldsymbol{\Sigma})$, we sample $\tilde{X}$ from $\mathcal{N}(\mu, \mathbf{V})$, where
$\mu=X-X \boldsymbol{\Sigma}^{-1} \operatorname{diag}(\mathbf{s})$
and $\mathbf{V}=2 \operatorname{diag}(\mathbf{s})-\operatorname{diag}(\mathbf{s}) \boldsymbol{\Sigma}^{-1} \operatorname{diag}(\mathbf{s})$, with $\operatorname{diag}(\mathbf{s})$ any diagonal matrix. If $\operatorname{diag}(\mathbf{s})$ is chosen such that $\mathbf{G}$ is positive semidefinite, we obtain valid knockoffs since:
$$
[X, \tilde{X}] \sim \mathcal{N}(0, \mathbf{G}) \quad \text{where} \quad
\mathbf{G}=\left(\begin{array}{cc}
\boldsymbol{\Sigma} & \boldsymbol{\Sigma}-\operatorname{diag}(\mathbf{s}) \\
\boldsymbol{\Sigma}-\operatorname{diag}(\mathbf{s}) & \boldsymbol{\Sigma}
\end{array}\right).
$$
In particular, this is the case with the equi-correlated construction where $s_j^{\mathrm{EQ}}=2 \lambda_{\min }(\boldsymbol{\Sigma}) \wedge \frac{1}{p}tr(\boldsymbol{\Sigma}) $ for all $j$.\\

In practice, even when $X$ can be assumed to be Gaussian, its covariance matrix $\boldsymbol{\Sigma}$ is unknown and must be estimated from the data.
This is generally done via shrinkage procedures such as Ledoit-Wolf shrinkage \citep{ledoit2003honey} or Graphical Lasso \citep{friedman2008sparse}. 
The validity of the resulting knockoffs directly depends on the accuracy of the covariance estimate.
Covariance estimation in high-dimensional regimes is a recognized as a difficult problem \citep{stein1972improving, fukunaga2013introduction}.
Indeed, this estimation requires to find an acceptable compromise between data fit and positive definiteness, along sparsity of the inverse covariance estimation from the Graphical Lasso. 
L2 shrinkage covariance estimation \citep{ledoit2003honey} on the other hand, is known to lead to excessive bias \citep{belilovsky2017learning}.
In addition to the intrinsic computation cost, hyperparameter setting for the Graphical Lasso is challenging, and leads to difficult and costly parameter selection.

\subsection{Existing nonparametric knockoff generation methods}
Recent machine learning techniques, such as deep learning, have also been used to capture complex data dependence structures in the context of knockoffs. \citet{romano2020deep} propose using deep neural networks to build knockoffs based on a target covariance matrix estimated from the data, as in the Gaussian algorithm. This approach represents a second-order approximation. The loss function also includes a term derived from the Maximum Mean Discrepancy to account for higher-order effects.  

Furthermore, to ensure a powerful procedure, the knockoff covariates should be as different as possible from the original ones. To achieve this, an additional term is incorporated to penalize high correlation between them.  

Using similar ideas, \citet{zhu2021deeplink} and \citet{liu2018auto} propose using a Variational Auto-Encoder (VAE) to build knockoffs as VAEs allow learning and sampling from complex data distributions. However, in the context of high-dimensional variable selection, the number of available samples may be insufficient to properly train a Deep Learning model. A limitation of these approaches is that they offer no theoretical guarantees on knockoffs exchangeability.

\subsection{Knockoff generation methods derived from the SCIP algorithm}
\label{sec:NGKO}
\subsubsection{The Sequential Conditional Independent Pairs algorithm}

\citealp{candes2018panning} have introduced a theoretical nonparametric knockoff generation approach known as the Sequential Conditional Independent Pairs (SCIP) algorithm.
It consists in progressively learning the distribution $X_j$ given $\mathbf{x}_{-j}$ and the previous knockoffs as described in Algorithm~\ref{algo:scip}.
\begin{algorithm}
\caption{Sequential Conditional Independent Pairs (SCIP; \citealp{candes2018panning})\label{algo:scip}}
$j=1$\\ 
\While{$j \leq p$}{
    Sample $\tilde{X}_j$ from $\mathcal{L}\left(X_j \mid X_{-j}, \tilde{X}_{1: j-1}\right)$\\
    $j=j+1$}
\end{algorithm}
This algorithm builds valid knockoffs assuming all the conditional distributions $\mathcal{L}\left(X_j \mid X_{-j}, \tilde{X}_{1: j-1}\right)$ to be known \citep{candes2018panning}.
However, as also noted by \cite{candes2018panning}, this knowledge is generally not available in practice.
Another caveat of this approach is its sequential nature: Algorithm~\ref{algo:scip} requires computing a new conditional distribution at each step without the possibility of parallelization, making computations potentially intractable.
These two practical limitations are discussed and addressed in the remainder of this section.\\

\subsubsection{Learning the conditional distributions}
\label{sec:learn-cond-dist}
To instantiate Algorithm~\ref{algo:scip}, we propose using a machine learning model $f\in \mathcal{F}$ to learn the conditional distributions.
A related idea is used for Variable Importance in the Conditional Permutation Importance of \citealp{chamma2023statistically} and proved valid in \citealp{lobo2025sobolcpidoublyrobustconditional}.
We observe from Algorithm \ref{algo:scip} that the order in which the knockoff variables are constructed is arbitrary.
Therefore, for simplicity, we fix the order to match the input order. 
For each $j$, $f_j$ predicts $X_j$ from $X_{-j}, \tilde{X}_{1: j-1}$.
Typically, using a Lasso model is a good default choice because, in the framework of \citealp{candes2018panning}, the relationship between the input covariates is assumed to be simple, and this choice helps to keep the algorithm computationally tractable.
In practice, we set $\lambda = \lambda_{max}/100$ with $\lambda_{max} = \frac{1}{n}\left\|(\mathbf{X}_{-j}, \mathbf{\tilde{X}}_{1: j-1})^T \mathbf{X}_j\right\|_{\infty}$. 
%
Then, the residual $\widehat{\boldsymbol{\epsilon}}_j = \mathbf{X}_j - f_j(\mathbf{X}_{-j}, \mathbf{\tilde{X}}_{1: j-1})$ is computed. 
The $j$-th knockoff variable is built by drawing a residual at random: $\tilde{\mathbf{X}}_j = f_j(\mathbf{X_{-j}, \tilde{X}_{1: j-1}}) + \sigma({\widehat{\boldsymbol{\epsilon}}_{j}})$, where $\sigma$ is a permutation of $\llbracket n \rrbracket$. 
The complete approach is described in Algorithm \ref{algo:seqgen}.\\

\begin{algorithm}[H]
\caption{Sequential generation of nonparametric knockoffs by learning to predict $\mathbf{X}_j$ from $(\mathbf{X_{-j}, \tilde{X}_{1: j-1}})$ using a model $f_j$. \label{algo:seqgen}}
\SetKwInOut{Input}{Input}
\Input{$f\in \mathcal{F}$}
\For{$j \in [1, p]$}{Fit a prediction model $f_j$ on $((\mathbf{X}_{-j}, \mathbf{\tilde{X}}_{1: j-1}), \mathbf{X}_j)$ \Comment{Typically a Lasso model}\\
Compute the residual $\widehat{\boldsymbol{\epsilon}}_j = \mathbf{X}_j - f_j((\mathbf{X}_{-j}, \mathbf{\tilde{X}}_{1: j-1}))$\\
$\text{Sample } \mathbf{\tilde{X}}_j = f_j((\mathbf{X}_{-j}, \mathbf{\tilde{X}}_{1: j-1})) + \sigma(\widehat{\boldsymbol{\epsilon}}_{j})$
}
{\bfseries Return} $\mathbf{\tilde{X}_{1: p}}$
\end{algorithm}

\medskip
We turn to showing that under mild assumptions, this indeed amounts to sampling from $\mathcal{L}\left(X_j \mid X_{-j}, \tilde{X}_{1: j-1}\right)$. 
In practice, we need to take into account the fact that $f^{*}_{j}\left(X_{-j}, \tilde{X}_{1: j-1}\right) := \e{X_j\mid X_{-j}, \tilde{X}_{1: j-1}}$ is estimated from a finite number of samples. To ensure asymptotic guarantees on the generated distribution, we require each regressor to be sufficiently accurate:

\begin{definition}[Estimator consistency]
    Given a training set of size $n$ used to regress an output $Y$ given an input $X$, a regressor $f_n$ is said to be consistent if $$\e{\left(f_n(X)-\e{Y|X}\right)^2}\xrightarrow[\enskip n \to \infty \enskip]{} 0.$$\label{def:consistency}
\end{definition}

In the following result, we establish the asymptotic validity the above sampling scheme: under the assumption of consistent regressors, the Wasserstein-2 distance between the samples produced by Algorithm~\ref{algo:seqgen} and the desired conditional distribution converges to 0 as the sample size grows to infinity.

\begin{proposition}[Validity of auto-regressive sampling]
For each $j\in \llbracket p \rrbracket$, assuming that $X$ is Gaussian and assuming  the consistency of each regressor $f_{j}\in \mathcal{F}$, with $\mathcal{F}$ a $P$-Donsker class, then $\tilde{X}_j|X_{-j}, \tilde{X}_{1:j-1}\sim \widehat{P}_j$ such that $\widehat{P}_j\xrightarrow[ n \to \infty]{W_2}\mathcal{L}\left(X_j \mid X_{-j}, \tilde{X}_{1: j-1}\right)$.\label{prop:empCondSamp}
\label{prop:valid-ar-sampling}
\end{proposition}

\subsubsection{Parallel knockoff generation}

In practice, Algorithm \ref{algo:seqgen} is computationally costly to run. Fitting the models $f_j$ cannot be parallelized as each fit depends on previously built knockoffs. 
To make computations tractable, we propose an approximate version of this Algorithm, in which samples are drawn from $\mathcal{L}\left(X_j \mid X_{-j}\right)$ instead of $\mathcal{L}\left(X_j \mid X_{-j}, \tilde{X}_{1: j-1}\right)$. In practice, this allows us to fit all models $f_j$ in parallel. Once this is done for all variables, knockoffs are built by shuffling the residuals:  $\mathbf{\tilde{X}}_j \text { is chosen as } f_j(\mathbf{X}_{-j}) + \sigma(\widehat{\boldsymbol{\epsilon}}_{j})$ with $\sigma$ a permutation of $\llbracket n \rrbracket$.
 
In words, we remove conditioning on previously built knockoff variables to make the algorithm easily parallelizable, as described in Algorithm \ref{algo-gen}.\\
\begin{algorithm}
\caption{Parallel generation of nonparametric knockoffs by learning to predict $X_j$ from $X_{-j}$ using a model $f_j$. \label{algo-gen}}
\SetKwInOut{Input}{Input}
\Input{$f$}
\For{$j \in [1, p]$}{Fit a prediction model $f_j$ on $(\mathbf{X}_{-j}, \mathbf{X}_j)$ \Comment{Typically a Lasso model}\\
Compute the residual $\widehat{\boldsymbol{\epsilon}}_j = \mathbf{X}_j - f_j(\mathbf{X}_{-j})$}

\For{$j \in [1, p]$}{$\text{Sample } \mathbf{\tilde{X}}_j = f_j(\mathbf{X_{-j}}) + \sigma({\widehat{\boldsymbol{\epsilon}}_{j}})$ \Comment{$\sigma$ is a permutation of $\llbracket n \rrbracket$}\\
}
{\bfseries Return} $\mathbf{\tilde{X}}_{1: p}$
\end{algorithm}
This approximation can have implications on the validity of the resulting knockoffs, which are studied in detail in Appendix \ref{appendix:parallel}.
In short, this approximation is reasonable in cases where there is a strong dependence structure in the data.
We show empirically that in high-dimensional problems, this approximation outputs results  very close to the theoretically grounded sequential approach.
Intuitively, the stronger the dependence structure, the more information about $X_j$ is contained in $X_{-j}$.\\

These nonparametric approaches circumvent the difficult problem of covariance estimation in high-dimensional settings. Note that Lasso models are usually of interest in these settings because, first, in the knockoffs setting, the input distribution $X$ is assumed to be simple, and the relationship between the input and the output $Y \mid X$ is the complex part (see \citealp{candes2018panning}). Moreover, assuming $X$ to be Gaussian, the relationship between covariates is linear. Additionally, it is a computationally efficient procedure. Finally, it fulfills the required complexity (see Example 19.7 from \citealp{van2000asymptotic}). Nevertheless, there may be cases in which a more complex class of functions is desired in order to deal with non-linearities, for instance, and therefore the Donsker assumption is not fulfilled. In these cases, similarly to what is proposed in the cross-fitted version in \citealp{williamson2021general}, a valid algorithm that takes residuals from a separate test set is possible (see Appendix \ref{sec:alg_test_train}).


\section{When knockoffs fail: diagnosing non-exchangeability}
\label{sec:KOfail}

\begin{figure}[h]
\centering
\begin{picture}(400,200) 
    \put(40,100){\textbf{Oracle knockoffs}} 
    \put(170,100){\textbf{Gaussian knockoffs}} 
    \put(295,100){\textbf{Parallel NP knockoffs}} 
    \put(0,0){\includegraphics[trim=0 0 0 27, clip, width=0.99\textwidth]{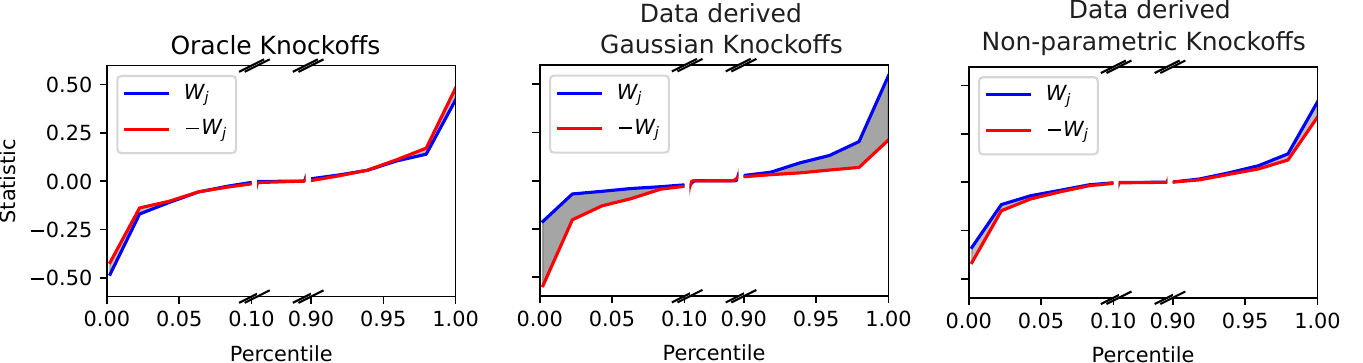}} 
\end{picture}
\caption{\textbf{Inverse CDFs of Knockoff statistics for null variables for oracle and data-derived Knockoffs.} Left panel: the empirical distribution for oracle knockoff statistics, obtained as Gaussian knockoffs with known covariance, is practically symmetric: this ensures the practical validity of the inference. Central panel: the empirical distribution of knockoff statistics in the same setup are built with a covariance estimate that relies on the Graphical Lasso, is not symmetric: indicating that real and knockoff importance scores are not comparable. Right panel: the nonparametric (NP) knockoff construction of Algorithm \ref{algo-gen} recovers the desired symmetry property, which ensures reliable inference. We use $n = 500$ samples and $p = 500$ variables.}
\label{fig:symmetryKO}
\end{figure}

A direct consequence of non-exchangeability is that knockoff statistics of null variables ${\{X_j \text{ for } j \in H_0\}}$ are no longer symmetrical -- yet this property is key to achieve error control via valid knockoffs as in \citealp{candes2018panning}.
Said otherwise, non-exchangeability can make knockoff importance scores non comparable with real variable importance scores which leads to bias in knockoff statistics. 
This is illustrated in Figure \ref{fig:symmetryKO} using a simulated data setup described in Section \ref{sec:results}.\\

On the left panel, we display the inverse Cumulative Distribution Function (CDF) of knockoff statistics of null variables, where Gaussian knockoffs come from an oracle, with known covariance. 
On the central panel, we display the inverse CDF in the same setup but with a covariance estimate that relies on the Graphical Lasso. On the right panel, we display the inverse CDF of statistics obtained using nonparametric parallel knockoffs built from the data using Algorithm~\ref{algo-gen}.
The knockoff statistics of the Oracle knockoffs are almost perfectly symmetric, as shown in the left panel. Note that knockoff statistics of null variables using data-derived Gaussian knockoffs are not symmetric: they are skewed toward positive values, signaling that real and knockoff importance scores are not comparable. Nonparametric knockoffs nearly restore the exact symmetry of the null knockoff statistics.

In the remainder of this section, we will focus on diagnosing this problem before assessing its consequences on error control in practice.\\

\textbf{A necessary condition for exchangeability.} 
Clearly, knockoffs exchangeability does not hold if $X$ is non-Gaussian while $\tilde{X}$ is Gaussian. One can simply take $S = \llbracket p \rrbracket$; then $(X, \tilde{X})_{swap(S)} =  (\tilde{X}, X) \stackrel{d}{\neq} (X, \tilde{X})$. 
More broadly, any knockoff generation procedure that relies on unfulfilled assumptions on variable distribution or dependence  structure fails to replicate the distribution of $X$. 
This leads to a natural sufficient condition to diagnose non-exchangeability: if there exists a classifier that is able to accurately distinguish samples from $X$ versus samples from $\tilde{X}$, then $\tilde{X} \neq X$ in distribution and exchangeability is violated. 
Classification has also been previously used in other contexts of Conditional Independence. Indeed, this idea is also related to the mimic-and-classify approach of \citet{sen2018mimicclassifymetaalgorithm}, inspired by Generative Adversarial Networks (GANs). In order to test conditional independence, this algorithm first tries to mimic the conditional independence distribution and then trains a discriminator to differentiate between the original and the mimicked distribution. In our case, this classifier is not trained to test conditional independence but rather as a prerequisite to enable testing conditional independence using knockoffs. This idea is also related to the C2ST (Classifier Two-Sample Testing) literature \citep{gretton2012kernel, lopez2017revisiting}.\\

\textbf{Classifier Two-Sample Testing for knockoffs.} Formally, we wish to test the null hypothesis $H_0: \tilde{X} \stackrel{d}{=} X$ given $n$ samples of each distribution. Following Section 3 of \citealp{lopez2017revisiting}, we proceed by constructing the dataset
$$
\mathcal{D}=\left\{\left(x^i, 0\right)\right\}_{i=1}^n \cup\left\{\left(\tilde{x}^i, 1\right)\right\}_{i=1}^n=:\left\{\left(z^i, l^i\right)\right\}_{i=1}^{2 n} .
$$

Then, the $2n$ samples of $\mathcal{D}$ are shuffled at random and split into disjoint training and testing subsets $\mathcal{D}_{\text {tr }}$ and $\mathcal{D}_{\text {te }}$, where $\mathcal{D}=\mathcal{D}_{\text {tr }} \cup \mathcal{D}_{\text {te }}$. Note that in practice, this split is performed several times to mitigate randomness, as in classical cross-validation.

\begin{figure}[H]
\centering
\includegraphics[width=0.75\linewidth]{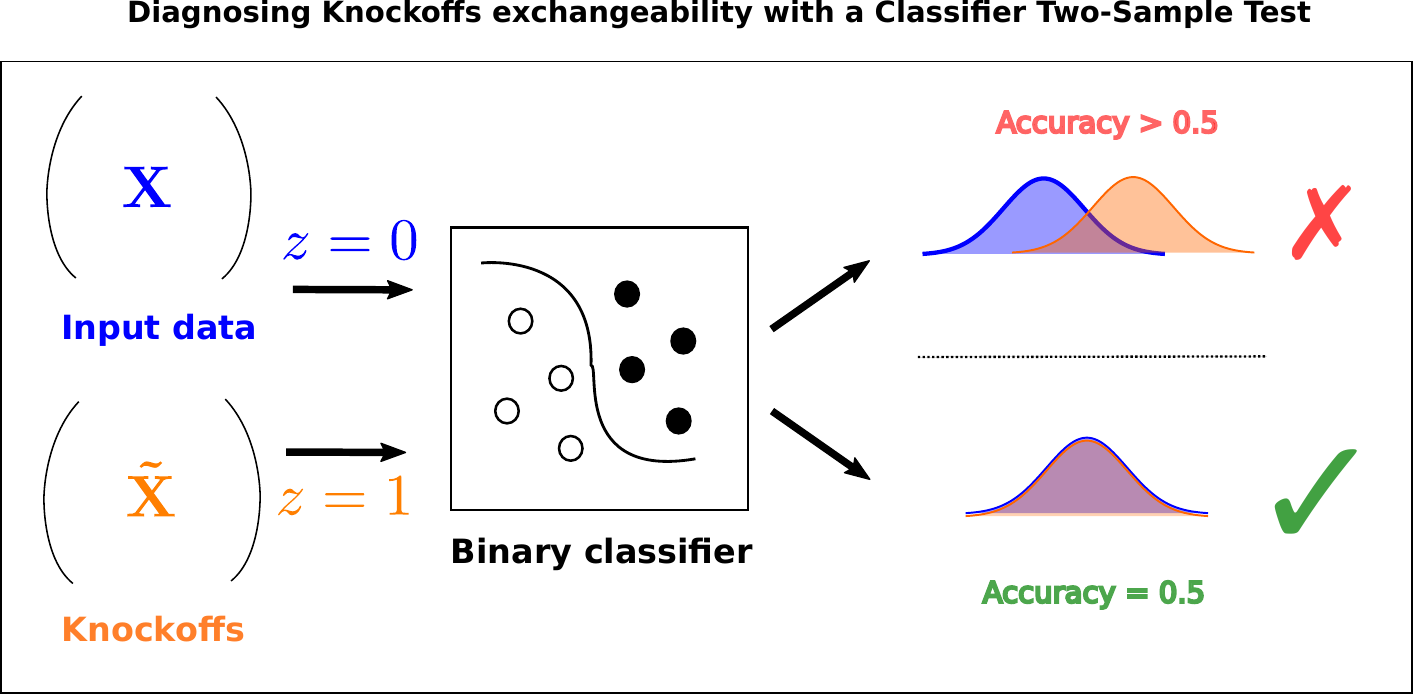}
\caption{ \textbf{Using a Classifier Two-Sample Test to diagnose exchangeability problems in knockoffs.} A binary classifier is fed withs two samples: a data set of interest $X$ and its associated knockoff candidates $\tilde{X}$. If the classifier's accuracy is above chance level, meaning that it can distinguish between real variables and knockoff variables, then the joint distributions of input data and knockoffs are not equal, indicating a violation of  exchangeability.}
\label{fig:c2stko}
\end{figure}

\begin{definition}[C2ST statistic, \citealp{lopez2017revisiting}]
  \label{def:c2st}

  Given $\mathcal{D}_{\text {tr }}$ and $\mathcal{D}_{\text {te }}$, where $\mathcal{D}=\mathcal{D}_{\text {tr }} \cup \mathcal{D}_{\text {te }}$ with $n_{\text {te }}:=\left|\mathcal{D}_{\text {te }}\right|$ and a binary classifier $g: \mathcal{R}^p \rightarrow\{0,1\}$ trained on $\mathcal{D}_{\text {tr}}$ the C2ST statistic $\hat{t}$ is defined as the classification accuracy on $\mathcal{D}_{\text {te }}$:

  $$
    \hat{t}=\frac{1}{n_{\text {te }}} \sum_{\left(z^i, l^i\right) \in \mathcal{D}_{\text {te }}} 1\left[g\left(z^i\right) = l^i\right].
    $$

\end{definition}
Intuitively, if $\tilde{X} \stackrel{d}{=} X$, the test accuracy $\hat{t}$ should remain near 0.5, corresponding to chance-level. 
Conversely, if $\tilde{X} \stackrel{d}{\neq} X$ and that the binary classifier is able to unveil distributional differences between the two samples, the test classification accuracy $\hat{t}$ should be greater than chance-level. 
The procedure is summarized in Figure \ref{fig:c2stko}.\\

\textbf{Exchangeability violation by improper sample pairing.} Note that the C2ST diagnostic tool only tests that knockoffs are indeed sampled from the same distribution as the original data. In practice, to have valid knockoffs, a consistent pairing between original and knockoffs samples is also needed. Said otherwise, knowing how to sample from the original data distribution is not enough to build valid knockoffs. We illustrate this point via a simple experiment in the same simulation setup described in Section \ref{sec:results}. The experiment consists in shuffling $50\%$ of the sample pairings of valid knockoffs. The middle panel reports the FDP for each simulation run in both scenarios (shuffled and non-shuffled). The left and right panels provide a toy illustration of shuffling sample pairings in $2D$. By construction, the C2ST diagnostic is invariant to this shuffling.

\begin{figure}[H]
\centering
\includegraphics[width=0.99\linewidth]{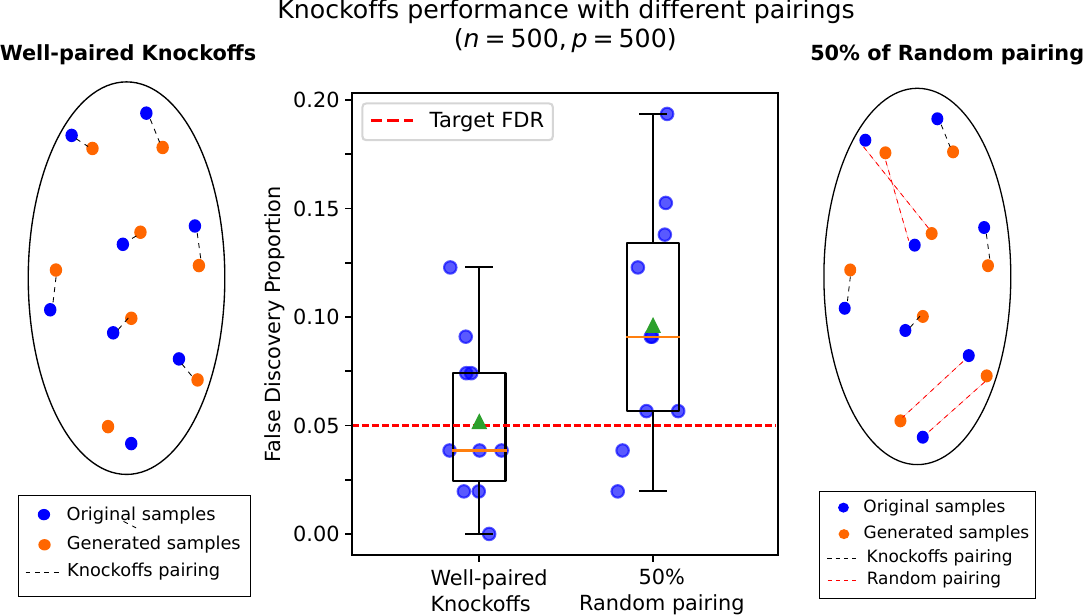}
\caption{ \textbf{Exchangeability violation by improper sample pairing.} In the left boxplot, we build valid knockoffs via the Gaussian algorithm and perform inference as in \citep{candes2018panning}. In the right boxplot, we use the same knockoffs but shuffle $50\%$ of the sample pairings before inference. We repeat this experiment $10$ times. Note that, coherently with theory, the FDR $(=\mathbb{E}(\mathrm{FDP}))$ is indeed controlled at the target level (5\%) in the left boxplot. However, in the right boxplot, error control is lost due to improper sample pairings. The left and right panels provide a toy illustration of shuffling sample pairings in $2D$.}
\label{fig:pairingsfailure}
\end{figure}

As seen in Figure \ref{fig:pairingsfailure}, FDR control is lost when shuffling $50\%$ of the sample pairings of valid knockoffs. This shows that, beyond equality of the distributions, a proper sample pairing is needed to obtain valid knockoffs. To test this in practice, one may compute the optimal assignment between real samples and knockoff samples e.g. via the Hungarian algorithm \citep{kuhn1955hungarian}. If the resulting assignment doesn't match the original one, then knockoffs cannot be exchangeable.

\section{Experimental results}
\label{sec:results}
The code for the proposed diagnostic tool and alternative nonparametric knockoffs algorithm is available at \url{https://github.com/alexblnn/KnockoffsDiagnostics}.
This section aims at illustrating some practical consequences of exchangeability violations on error control and diagnosis via C2ST statistics. 
\subsection{Simulated data experiments}

To assess the consequences of varying degrees of non-exchangeability with Gaussian knockoffs, we design a simulation setup in which we control the difficulty to produce valid knockoffs. 
At each simulation run, we first generate i.i.d. Gaussian data with a tensor structure $\mathbf{X} \in \mathbb{R}^{n \times a \times b \times c}$ with $a \times b \times c = p$. The idea is to generate $n$ samples in a 3D discrete feature space with shape $(a, b, c)$.
Then, we apply an isotropic smoothing kernel of width $w$ across these three dimensions, mimicking a smooth three-dimensional structure, and flatten the data to obtain $\mathbf{X} \in \mathbb{R}^{n \times p}$.\\

Then, we draw the true support $\boldsymbol{\beta}^* \in \{0, 1\}^p$. The number of non-null coefficients of $\boldsymbol{\beta}^*$ is controlled by a sparsity parameter $s_p$, i.e. $s_p = \lVert \boldsymbol{\beta}^* \rVert_0/p$. The target variable $\mathbf{y}$ is built
using a linear model:
\begin{align}\mathbf{Y} = \mathbf{X}\boldsymbol{\beta}^* + \sigma\boldsymbol{\epsilon},\nonumber\end{align}
where $\sigma = \lVert \mathbf{X} \boldsymbol{\beta}^* \rVert_2 /
(\text{SNR}\lVert\boldsymbol{\epsilon}\rVert_2)$ controls the amplitude of the noise, SNR being the signal-to-noise ratio.
We choose the setting $n = 500, p = 500, s_p = 0.1, \text{SNR} = 2$. To obtain $p = 500$, we use $(a, b ,c) = (10, 10, 5)$. Note that using $w = 0$ is equivalent to sampling i.i.d. Gaussians since no smoothing is applied in this case.
We vary the kernel width  $w$ in the interval $[0, 1.25]$ to parametrize the difficulty of the problem. This parameter is monotonically related to the level of correlation observed in the data; as the kernel width increases, the data becomes increasingly correlated.\\

In all settings, we use the Graphical Lasso for covariance estimation \citep{friedman2008sparse}. For each of $N$ simulations, we compute the empirical FDP of the vanilla knockoff selection set $S$ and the C2ST statistic averaged across 5-fold cross-validation: \begin{align}\widehat{\mathrm{FDP}}(S) = \frac{|S\cap \mathcal{H}_0|}{|S|},\quad \hat{t}=\frac{1}{n_{\text {te }}} \sum_{\left(z^i, l^i\right) \in \mathcal{D}_{\text {te }}} 1\left[g\left(z^i\right) = l^i\right].\nonumber\end{align}

\subsubsection{Varying number of samples and variables}

We first benchmark all existing methods for building knockoffs for continuous data enumerated in Section \ref{sec:genKO} using $w = 0.5$. This value yields local correlations levels that are comparable to those observed in fMRI data. VAE was implemented using \url{https://github.com/zifanzhu/DeepLINK} and Deep Knockoffs were implemented using \url{https://github.com/msesia/deepknockoffs}.

\begin{figure}[H]
\centering
\includegraphics[width=0.47\linewidth]{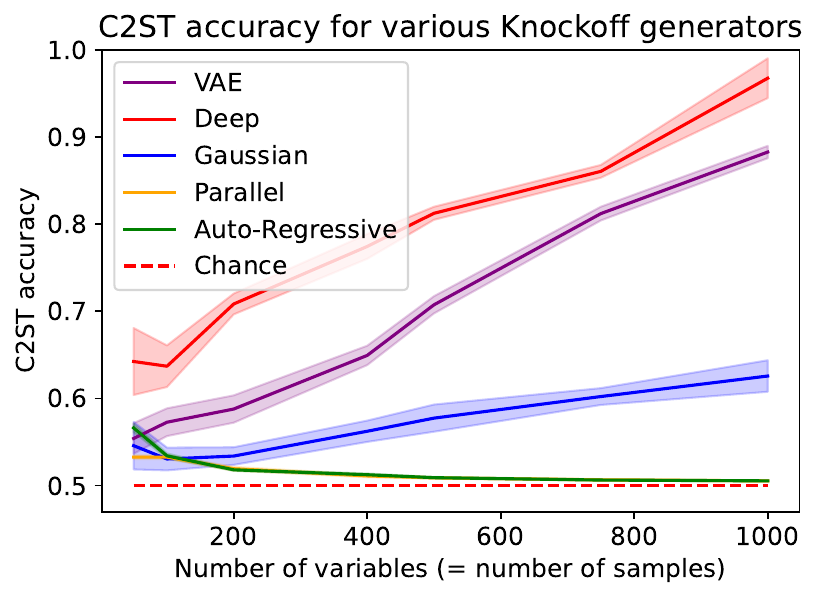}
\includegraphics[width=0.51\linewidth]{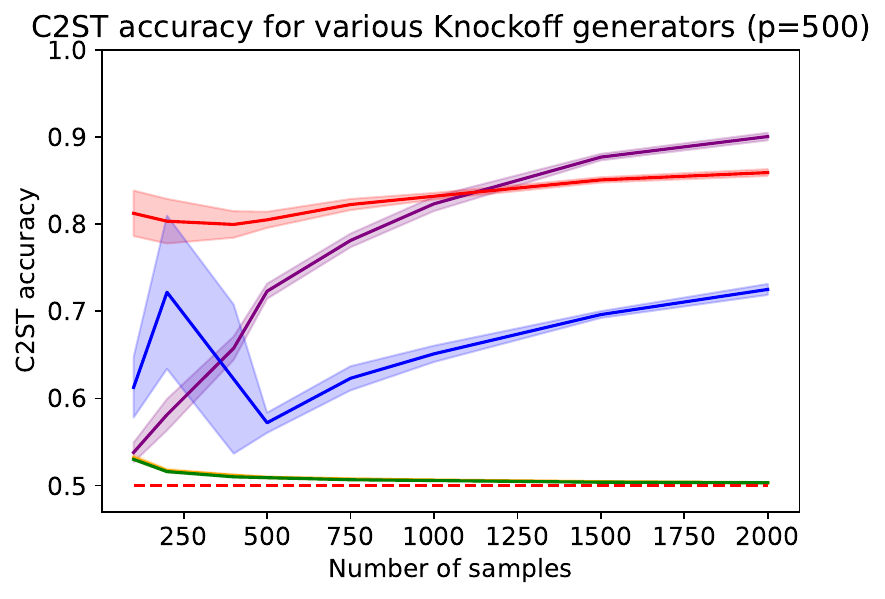}
\caption{\textbf{C2ST diagnostic metric for varying number of variables and samples.} In the left panel, the number of variables $p$ grows and $n = p$ at each point. Note that, when $p > 100$ the C2ST accuracy is clearly above chance level for all methods apart from Gaussian knockoffs and nonparametric knockoffs, signaling non-exchangeability in this regime. Parallel and auto-regressive nonparametric knockoffs yield very similar results. In the right panel, the number of variables is kept constant with $p$ set to $500$. The number of samples varies from $100$ to $2000$. Apart from nonparametric (parallel or auto-regressive) knockoffs, all other methods fail to produce valid knockoffs, even for large values of $n$.}
\label{fig:relatedwork}
\end{figure}

The left panel of Figure \ref{fig:relatedwork} shows that all methods except Gaussian and nonparametric knockoffs fail when $p$ grows larger (using $n=p$). Amongst the state-of-the-art methods, Gaussian knockoffs are the best candidate as they maintain a C2ST accuracy of at most $0.6$. Nonparametric knockoffs maintains chance-level accuracy in all settings. Note that the data at hand is Gaussian -- therefore, the non-exchangeability of Gaussian knockoffs is necessarily due to the problem of covariance estimation. In contrast, the proposed nonparametric approach circumvents this difficult estimation problem by relying on regressions.\\

The right panel of Figure \ref{fig:relatedwork} shows that increasing the number of samples $n$ for a fixed $p$ does not fix the problem, as all methods apart from nonparametric (parallel or auto-regressive) knockoffs fail to provide C2ST-valid knockoffs. C2ST accuracy rises as $n$ grows larger for all methods apart from nonparametric knockoffs. Intuitively, one might expect that as $n$ grows larger, producing valid knockoffs becomes easier and therefore that the C2ST accuracy should decrease. A possible interpretation of this result is the presence of two competing effects: while learning features of the underlying joint distribution of the data is indeed easier when $n$ grows, the number of training samples accessible to the classifier also grows as it is equal to $2n$. Therefore, the discriminating power of the classifier also improves with larger values of $n$. From now on, we discard the VAE and Deep Learning approach because of their poor empirical performance.\\

\subsubsection{Varying smoothing}

We now turn to evaluating the performance of Gaussian and nonparametric knockoffs for varying smoothing. We use the 3D smoothing kernel width to parametrize the correlation present in the data, which controls the difficulty to produce valid knockoffs.

\begin{figure}[H]
\centering
\includegraphics[width=0.99\linewidth]{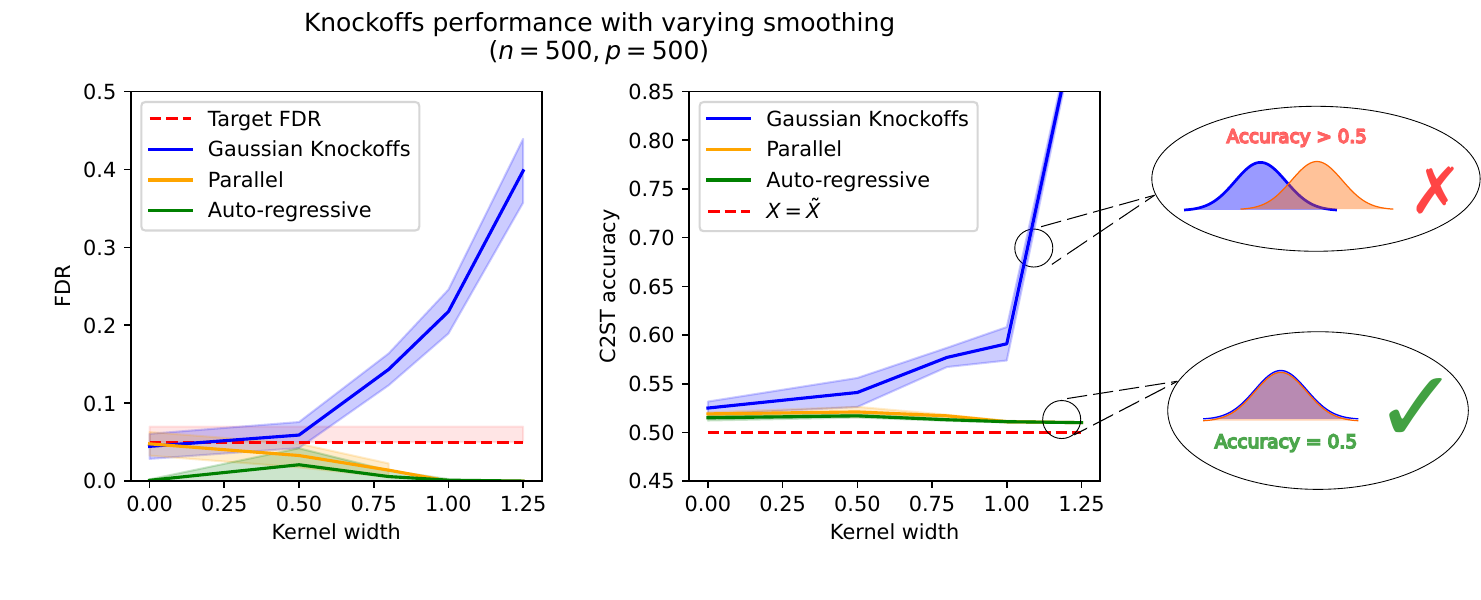}
\caption{ \textbf{FDP and C2ST diagnostic metric for varying smoothing.} We use the 3D smoothing kernel width to parametrize the correlation present in the data, which in turn tunes the difficulty to produce valid knockoffs. For Gaussian knockoffs, the FDR is controlled only in the easiest settings, i.e. $w \in [0, 0.5]$. Note that using $w = 0.5$ represents a common setting encountered on real data: it indeed yields local correlations levels that are comparable to those observed in fMRI data. For $w > 0.5$ the achieved FDR is substantially above the target FDR, and it grows as the smoothing increases.
Note that the C2ST accuracy is clearly above chance level for $w > 0.5$, signaling non-exchangeability in this regime. Nonparametric knockoffs -- defined in Section \ref{sec:NGKO} -- preserve error control in all regimes, which is consistent with C2ST accuracy remaining near chance level.}
\label{fig:diagsimu}
\end{figure}
The left panel of Figure \ref{fig:diagsimu} shows that, for Gaussian knockoffs, the FDR is controlled only in the easiest settings, i.e. $w \in [0, 0.5]$. For $w > 0.5$ the achieved FDR is substantially above the target FDR, and it grows as the smoothing increases. The $C2ST$ accuracy is clearly above chance level for $w > 0.5$. 
By contrast, nonparametric knockoffs maintain FDR control in all settings, while the $C2ST$ accuracy remains near chance level.\\

\subsection{Semi-simulated data setup.} We now evaluate knockoff generation methods and our proposed diagnostic tool with real data. Following \citealp{blain2023false, Nguyen2022CRT}, we use semi-simulated data to evaluate the proposed method with observed $\mathbf{X}$. We use a simulated response $\mathbf{Y}$ to be able to compute the FDP.
We consider a first functional Magnetic Resonance Imaging (fMRI) dataset $(\mathbf{X}_1, \mathbf{Y}_1)$ on which we perform inference using a Lasso estimator; this yields $\boldsymbol{\beta}_1^* \in \mathbb{R}^{p}$ that we will use as our ground truth. Then, we consider a separate fMRI dataset $(\mathbf{X}_2, \mathbf{Y}_2)$ for data generation. 
The point of using a separate dataset is to avoid circularity between the ground truth definition and the inference procedure. 
Concretely, we discard the original response vector $\mathbf{Y}_2$ for this dataset and build a simulated response $\mathbf{Y}_2^{sim}$ using a linear model:

\begin{align}\mathbf{Y}_2^{sim} = \mathbf{X}_2\boldsymbol{\beta}_1^* + \sigma\boldsymbol{\epsilon},\nonumber\end{align}

where we set $\sigma$ so that $SNR = 4$. We consider $7$ binary classifications problems like "gambling" (rewards vs loss) taken from the HCP dataset. For each of these classification problems, the dataset consists in  $778 \times 2 = 1556$ samples and $1000$ features. These features are obtained by averaging fMRI signals within a Ward parcellation scheme, which is known to yield spatially homogeneous regions \citep{thirion:ward}.\\

\begin{figure}[H]
\centering
\includegraphics[width=0.99\linewidth]{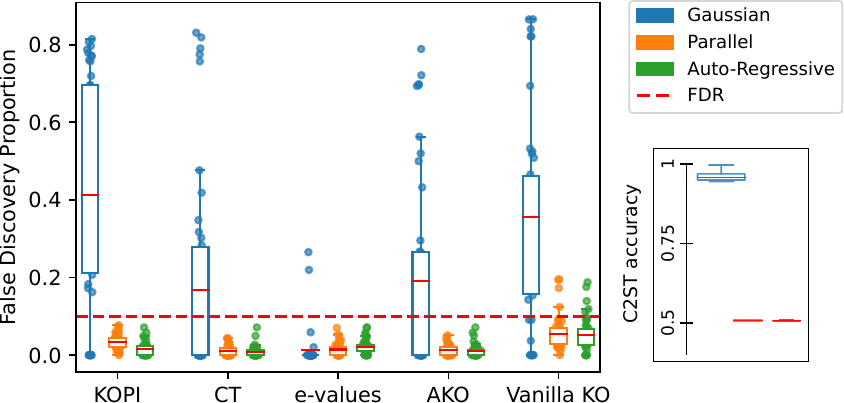}
\caption{\textbf{Empirical FDP on semi-simulated data for 42 contrast pairs using Gaussian vs nonparametric Knockoffs.} We use 7 HCP contrasts C0: "Motor Hand", C1: "Motor Foot", C2: "Gambling", C3: "Relational", C4: "Emotion", C5: "Social", C6: "Working Memory". We consider all 42 possible train/test pairs: the train contrast is used to obtain a ground truth, while the test contrast is used to generate the response. Inference is performed using the 5 methods considered in the paper and the empirical FDP is reported. Notice that error control of all methods is violated using default Gaussian knockoffs (left panel) and recovered using nonparametric knockoffs (right panel). The C2ST metric is coherent in both cases, with an average accuracy of $0.95$ for the Gaussian knockoffs (signalling non-exchangeability) and $0.51$ for the nonparametric knockoffs.
}%
\label{fig:nocontrolvsNG}
\end{figure}

As we use $7$ datasets of HCP, we obtain $42 = 7 \times 6$ possible pairs. Since we consider $\boldsymbol{\beta}_1^*$ as the ground truth, the FDP can be computed. Figure \ref{fig:nocontrolvsNG} shows False Discovery Proportion levels for 42 semi-simulated fMRI datasets based on HCP data for 5 different knockoff-based inference methods, using either Gaussian knockoffs or nonparametric knockoffs. 
All methods exhibit problematic False Discoveries Proportions using default Gaussian knockoffs (left panel) and Graphical Lasso-based covariance learning. We refer to \citealp{blain2023false} for a more detailed comparison. The expected behavior is recovered using nonparametric knockoffs (right panel). 
As one could expect, the C2ST metric obtains high discrimination power ($0.95$ prediction accuracy) for the Gaussian knockoffs (signalling non-exchangeability), but is at chance (accuracy of $0.51$) for nonparametric knockoffs. Regarding the pairing condition, the Hungarian algorithm signals that Gaussian and nonparametric (Parallel as well as Auto-Regressive) knockoffs are optimally paired with original variables.

\subsection{Computational cost}

Computation time is benchmarked in Appendix \ref{sec:comptime}. As expected, the Parallel generation scheme is much more efficient than Auto-regressive Knockoffs, especially when a large number of CPUs can be used.
The parallel algorithm is also less costly in terms of computing time than the Gaussian algorithm. Using a single CPU, a $3\times$ speedup is achieved compared to the original algorithm. When leveraging 40 parallel CPUs, a $15\times$ speedup is achieved on fMRI data and other large problems.

\section{Discussion}

Knockoffs are a powerful and efficient method for controlled variable selection. This inference procedure allows performing conditional variable selection in one round of inference, without considering each variable individually as in Conditional Randomization Tests approaches.
The statistical guarantees provided by this method rely on the ability to construct valid knockoffs - namely, knockoffs and original variables must be exchangeable to ensure the control of false positives. However, constructing valid knockoffs based only on the available observations is not trivial.

The difficulty of this problem is particularly salient in contexts where \textit{i)} the data at hand has a strong dependence structure and \textit{ii)} the number of variables $p$ is large. These are common characteristics of high-dimensional variable selection problems where knockoffs are relevant, such as fMRI brain mapping or genomic data analyses. These characteristics make the estimation of the joint distribution of the data increasingly difficult, whether done through covariance estimation (as in the Gaussian algorithm) or via deep learning based techniques. We have shown that several existing knockoff generation procedures are unable to produce valid knockoffs in such settings.\\

This paper also provides insight into the consequences of exchangeability violations. Experiments on both real and simulated data highlight the substantial impact of deviations from exchangeability on the reliability of knockoff-based methods. These issues can lead to instability in error control for all knockoff-based approaches. This result is important because it contrasts with previous positive findings in the literature.  

On the one hand, \citet{fan2025asymptoticfdrcontrolmodelx} provided a general framework for achieving asymptotic FDR guarantees. However, this control does not hold in finite samples, as their asymptotic approximate symmetry assumption for the estimated $W$ is not satisfied, as shown in Figure \ref{fig:symmetryKO}. On the other hand, in model-X methods introduced by \citet{candes2018panning}, such as Conditional Randomization Tests (CRT) and knockoffs, the main assumption relies on knowing the input distribution in order to generate conditionally independent samples and knockoff covariates, respectively. These methods also depend on the precision of the test statistic to capture the relationship between input and output covariates, which determines statistical power.  

Nevertheless, the input distribution is generally unknown and must be estimated. In this context, an important question arises: do statistical guarantees still hold when this distribution is estimated? This question can be framed in terms of a \textit{double robustness} property, meaning that errors in estimating the input distribution could be compensated by a sufficiently accurate estimation of the input-output relationship.  

For the CRT, the Maxway CRT proposed by \citet{maxway} was introduced as a robust approach to input distribution estimation. However, \citet{DR_dcrt} showed that the distilled CRT (dCRT) from \citet{dcrt} was already double-robust when using partial correlations of the residuals as a statistic. Specifically, under certain convergence rate conditions for the estimates, the dCRT converges to the Generalized Covariance Measure (GCM) of \citet{Shah_2020}, which is also double-robust.  

For knockoffs, \citet{candes2018panning} conjectured a similar double robustness property, suggesting that when the covariance matrix is not well estimated, the FDR remains controlled at the cost of reduced power. In their experiments, where the covariates exhibit low correlation, FDR control is observed when the empirical covariance matrix is used. However, this comes at the expense of complete power loss, as no discoveries are made.  
Our simulations demonstrate that this property is not guaranteed \textit{a priori} for knockoffs in more realistic settings.  \\

In this paper, we have introduced a diagnostic tool that provides practitioners with a means to identify exchangeability problems. 
This tool relies on two parts: a classifier two-sample test that aims to detect distributional differences between original and knockoff variables, and a procedure that checks for proper pairing between original observations and their knockoff. Thus, we have two necessary (but not sufficient) conditions for exchangeability violation. 

As part of our efforts to mitigate the exchangeability problems associated with non-Gaussian data or poor covariance estimation, we have proposed an efficient alternative approach for constructing nonparametric knockoffs. Experiments on simulated data show that the knockoffs we obtain are C2ST-valid. This parallel approach yields similar results to the autoregressive approach for a fraction of the computational time.

\vfill
\newpage
\bibliographystyle{apalike}
\bibliography{bib.bib}       

\newpage
\begin{appendix}

  \section{Proof of Proposition \ref{prop:valid-ar-sampling}}
  \label{sec:proofs}
  The construction in Algorithm~\ref{algo:seqgen} is motivated by the following lemma, which states that when using the information from all covariates to predict another in a Gaussian vector, and adding noise from another sample, the result follows the conditional distribution and remains a Gaussian vector.
  
  \begin{lemma}[Gaussian conditional sampling] Let $X, X'$ two i.i.d. Gaussian vectors and $j\in\llbracket p \rrbracket$. Denote $\tilde{X}_j=\e{X_j|X_{-j}}+\left(X'_j-\e{X'_{j}|X'_{-j}}\right)$. Then $\tilde{X}_j|X_{-j}\sim \mathcal{L}(X_j|X_{-j})$. Moreover, $(X, \tilde{X}_j)$ is still a Gaussian vector. \label{lemma:GaussianCondSamp}
  \end{lemma}

  \begin{proof}[Proof of Lemma~\ref{lemma:GaussianCondSamp}]
Recall that in the Gaussian setting, we have $X_j|X_{-j} \sim \mathcal{N}(\mu_\mathrm{cond}^j, \Sigma_\mathrm{cond}^j)$ with $\mu_\mathrm{cond}^j=\mu_{j}+\boldsymbol{\Sigma}_{j, -j}\boldsymbol{\Sigma}^{-1}_{-j, -j}(X_{-j}-\mathbf{\mu}_{-j})$ and $\boldsymbol{\Sigma}_\mathrm{cond}^j=\boldsymbol{\Sigma}_{j, j}-\boldsymbol{\Sigma}_{j,-j}\boldsymbol{\Sigma}^{-1}_{-j, -j}\boldsymbol{\Sigma}_{-j, j}$. For simplicity of notation, we omit the dependence of these conditional parameters on $j$.

By definition $\e{X_j|X_{-j}}=\mu_\mathrm{cond}$. Observe that:
\begin{align*}
    X'_j-\e{X'_j|X'_{-j}}&=X'_j-\mu_j-\boldsymbol{\Sigma}_{j,-j}\boldsymbol{\Sigma}^{-1}_{-j,-j}(X'_{-j}-\mathbf{\mu}_{-j}).
\end{align*}
We easily observe that this is a centered Gaussian vector. We turn to deriving its variance:
\begin{align*}
    \mathbb{V}&(X'_j-\mu_j-\boldsymbol{\Sigma}_{j,-j}\boldsymbol{\Sigma}^{-1}_{-j,-j}(X'_{-j}-\mathbf{\mu}_{-j}))\\&= \mathbb{V}(X'_j)+\boldsymbol{\Sigma}_{j,-j}\boldsymbol{\Sigma}^{-1}_{-j,-j}\mathbb{V}(X'_{-j})\boldsymbol{\Sigma}^{-1}_{-j,-j}\boldsymbol{\Sigma}_{-j,j}-2\e{(X'_{j}-\mu_j)\boldsymbol{\Sigma}_{j,-j}\boldsymbol{\Sigma}^{-1}_{-j,-j}(X'_{-j}-\mathbf{\mu}_{-j})}\\
    &=\boldsymbol{\Sigma}_{j,j}+\boldsymbol{\Sigma}_{j,-j}\boldsymbol{\Sigma}^{-1}_{-j,-j}\boldsymbol{\Sigma}_{-j,j}-2\boldsymbol{\Sigma}_{j,-j}\boldsymbol{\Sigma}^{-1}_{-j,-j}\boldsymbol{\Sigma}_{-j,j}
    \\&=\boldsymbol{\Sigma}_{j,j}-\boldsymbol{\Sigma}_{j,-j}\boldsymbol{\Sigma}^{-1}_{-j,-j}\boldsymbol{\Sigma}_{-j,j}.
\end{align*}
We observe that this is exactly $\Sigma_\mathrm{cond}$ and therefore the result follows: $$\tilde{X}_{j}\mid X_{-j}=\mu_\mathrm{cond}+(X'_j-\e{X'_j|X'_{-j}})\sim \mathcal{N}(\mu_\mathrm{cond}, \boldsymbol{\Sigma}_\mathrm{cond}).$$

Finally, we observe that the added residual consists of a linear combination of the coordinates of a Gaussian vector, and is therefore Gaussian. Moreover, it is independent of $X$. Consequently, the joint vector is Gaussian. Thus, we observe that $(X, \tilde{X}_j)$ is a linear combination of this Gaussian vector, and is therefore also Gaussian.
\end{proof}

  \begin{proof}[Proof of Proposition~\ref{prop:valid-ar-sampling}]
Recall the 2-Wasserstein distance given by 
$$\left(\underset{P_\theta\in \Theta (\mu, \nu)}{\mathrm{inf}}\int \|x-y\|^2\mathrm{d}P_\theta (\mathrm{dx,dy})\right)^{\frac{1}{2}},$$
where $\Theta(\mu, \nu)$ is the set of distributions with marginals $\mu$ and $\nu$.

This result is done by induction. First, for the first coordinate, let's denote $P_1:=\mathcal{L}(X_1|X_{-1})$. Under the Gaussian assumption, from Lemma \ref{lemma:GaussianCondSamp} we have that $P_1\overset{d}{=}\e{X_1\middle| X_{-1}}+\left(X'_{1}-\e{X'_1|\mathbf{x}'_{-1}}\right)$. Let's also denote by $\widehat{P}_1:=f_{1}\left(X_{-1}\right)+X'_{1}-f_{1}\left(X'_{-1}\right)$ its empirical counterpart.  The goal is to show that $ W_2(P_1, \widehat{P}_1) \xrightarrow[\enskip n \to \infty \enskip]{} 0$:
    
\begin{align*}
    W_2(P_1, \widehat{P}_1) &= \left(
        \underset{P_\theta \in \Theta(P_1, \widehat{P}_1)}{\mathrm{inf}}
        \int_{\mathbb{R}^{2p} \times \mathbb{R}^{2p}} 
        (x-y)^2 P_\theta(\mathrm{dx, dy})
    \right)^{\frac{1}{2}} \\
    &\leq \left(
        \int_{\mathbb{R}^{2p}}
        \bigg(
            \mathbb{E}\left[X_1 \middle| X_{-1}\right] +
            \left(X'_1 - \mathbb{E}\left[X'_1 \middle| X'_{-1}\right]\right) \right. \\
            &\quad \left. - f_{1}\left(X_{-1}\right) -
            \left(X'_1 - f_{1}\left(X'_{-1}\right)\right)
        \bigg)^2
        P_{X}(\mathrm{dx}) 
        P_{X'}(\mathrm{dx}')
    \right)^{\frac{1}{2}} \\
    &= \left(
        \int_{\mathbb{R}^{2(p-1)}}
        \bigg(
            f^{*}_{1}\left(X_{-1}\right) -
            f^{*}_{1}\left(X'_{-1}\right) \right. \\
            &\quad \left. - f_{1}\left(X_{-1}\right) +f_{1}\left(X'_{-1}\right)
        \bigg)^2
        P_{X_{-1}}(\mathrm{dx}_{-1}) 
        P_{X'_{-1}}(\mathrm{dx}'_{-1})
    \right)^{\frac{1}{2}} \\
    &\leq 
    \left(
        \mathbb{E}\left[
            \left(
                f^{*}_{1}\left(\mathbf{x}_{-1}\right) - 
                f_{-1}\left(\mathbf{x}_{-1}\right)
            \right)^2
        \right]
    \right)^{\frac{1}{2}} \\
    &\quad +
    \left(
        \mathbb{E}\left[
            \left(
                f^{*}_{1}\left(X'_{-1}\right) - 
                f_{1}\left(X'_{-1}\right)
            \right)^2
        \right]
    \right)^{\frac{1}{2}}.
\end{align*}
To conclude, we note that there is a need for uniform control. This is achieved by the limited complexity of the class of regressors (Donsker class) and its consistency. Under these assumptions, we apply Lemma 19.24 from \citealp{van2000asymptotic}. Therefore, there is uniform convergence, and the last two terms vanish.

For the $j$-th coordinate, we proceed similarly as before. By the induction hypothesis we have that $\tilde{X}_l|X_{-l}, \tilde{X}_{1:l-1}\sim \mathcal{L}(X_l|X_{-l}, \tilde{X}_{1:l-1})$ for each $l<j$. Therefore, given $(X',\tilde{X}'_{1:j-1})\overset{\mathrm{i.i.d}}{\sim}(X,\tilde{X}_{1:j-1})$, let $P_j$ be the distribution of $\e{X_j\middle|X_{-j}, \tilde{X}_{1:j-1}}+\left(X'_{j}-\e{X'_j|X'_{-j}, \tilde{X}'_{1:j-1}}\right)$ conditionally on $(X_{-j}, \tilde{X}_{1:j-1})$. Analogously, let's denote its empirical counterpart as $\widehat{P}_j:=f_{j}\left(X_{-j}, \tilde{X}_{1:j-1}\right)+X'_{j}-f_{j}\left(X'_{-j}, \tilde{X}'_{1:j-1}\right)$. By Lemma \ref{lemma:GaussianCondSamp}, we have that $P_j$ is indeed the desired distribution, i.e. $\mathcal{L}(X_j|X_{-j}, \tilde{X}_{1:j-1})$. Similarly as before, using the consistency of $f_j$ we conclude that $ W_2(P_j, \widehat{P}_j) \xrightarrow[\enskip n \to \infty \enskip]{} 0.$
\end{proof}

\section{Nonparametric knockoffs with complex learners}\label{sec:alg_test_train}

Let's denote $\mathbf{X}^i$ as the $i$-th row of the matrix $\mathbf{X}$, which corresponds to the $i$-th individual.

\begin{algorithm}[H]
\caption{Cross-fitted sequential generation of nonparametric knockoffs by learning to predict $\mathbf{X}_j$ from $(\mathbf{X}_{-j}, \tilde{\mathbf{X}}_{1: j-1})$ using a model $f_j$. \label{algo:seqgen_cf}}
\SetKwInOut{Input}{Input}
\Input{$f, K$}
Generate $B_n\in [1, K]^n$ by sampling uniformly from $[1, K]$ with replacement, and for $k\in [1, K]$, denote by $D_k:=\{i\mid B_{n,i}=k\}$ the set of indices in the $k$-th fold 
\For{$k\in [1, K]$}{
\For{$j \in [1, p]$}{Fit a prediction model $f_{j, k}$ on $\{((\mathbf{X}^i_{-j}, \bar{\mathbf{X}}^i_{1: j-1}), \mathbf{X}_j^i)\}_{i \in \cup_{l\neq k}D_l}$ \\
Compute the residual $\widehat{\boldsymbol{\epsilon}}_{j, k} = \{\mathbf{X}^i_j - f_{j,k}((\mathbf{X}^i_{-j}, \tilde{\mathbf{X}}^i_{1: j-1}))\}_{i\in D_k}$\\
$\text{Sample } \bar{\mathbf{X}}^i_j = f_{j, k}((\mathbf{X}^i_{-j}, \bar{\mathbf{X}}^i_{1: j-1})) + \varepsilon_{i,j,k}$ with $\varepsilon_{i,j,k}$ sampled uniformly in $\widehat{\boldsymbol{\epsilon}}_{j, k}$ for $i\in \cup_{l\neq k}D_l$\\
$\text{Sample } \tilde{\mathbf{X}}^i_j = f_{j,k}((\mathbf{X}^i_{-j}, \tilde{\mathbf{X}}^i_{1: j-1})) +\varepsilon_{i,j,k}$ with $\varepsilon_{i,j,k}$ sampled uniformly in $\widehat{\boldsymbol{\epsilon}}_{j, k}$ for $i\in D_k$
}
}

{\bfseries Return} $\mathbf{\tilde{X}}_{1: p}$
\end{algorithm}

The proof of the validity of Algorithm \ref{algo:seqgen_cf} follows exactly the same reasoning as the one given in Appendix \ref{sec:proofs}, but it directly uses the consistency of the estimator to conclude, without requiring the uniform control needed to account for the dependency of the residuals on the trained regressors.

\section{Computation time of nonparametric knockoffs}
\label{sec:comptime}

\begin{figure}
\centering
\includegraphics[width=0.99\linewidth]{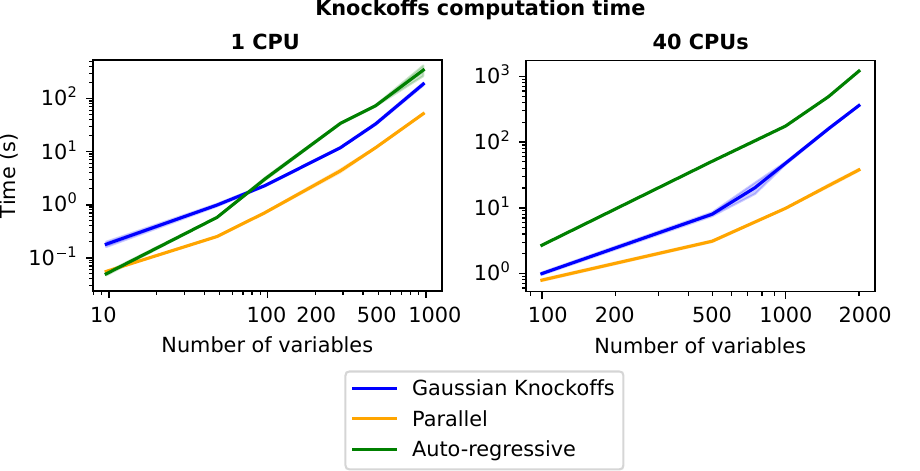}
\caption{\textbf{Nonparametric and Gaussian knockoffs computation time.} We use either one CPU or 40 CPUs and report the computation time for various problem dimensions in log-log scale. Note that using nonparametric (Parallel) Knockoffs on a single CPU yields around a $3\times$ speedup across all problem dimensions compared to Gaussian knockoffs. Using 40 CPUs, a $5\times$ speedup is achieved for problems under $1000$ variables, and a $10\times$ to $15\times$ speedup for problems of up to $2000$ variables.
}%
\label{fig:benchmarktime}
\end{figure}

We have performed benchmarks to evaluate the computation time needed for the proposed nonparametric knockoffs algorithm compared to the Gaussian algorithm. We use the simulation setup described in Section \ref{sec:KOfail} and vary the number of variables $p$. For all values of $p$, we use $n = p$ samples. We first consider using a single CPU and then consider using 40 CPUs. For the Gaussian approach, parallelization is done in the Graphical Lasso covariance estimator via the $n_{jobs}$ argument in scikit-learn \citep{pedregosa2011scikit}. Note that profiling the code when using a single CPU shows that most of the computation time is spent in covariance estimation. For the nonparametric approach, the 40 CPUs are used to train Lasso models that predict $X_j$ from $\mathbf{X_{-j}}$ in parallel. The result is shown in Fig. \ref{fig:benchmarktime}\\

As expected, the auto-regressive procedure is much slower than both other approaches: computing knockoffs is nearly $10\times$ slower than using the parallel procedure on 40 CPUs for 1000 variables. Using parallel nonparametric knockoffs on a single CPU yields around a $3\times$ speedup across all problem dimensions compared to Gaussian knockoffs. Using 40 CPUs, a $5\times$ speedup is achieved for problems under $1000$ variables, and a $10\times$ to $15\times$ speedup for problems of up to $2000$ variables.

\section{Theoretical analysis of parallel knockoff generation}
\label{appendix:parallel}

We start of by showing that the parallel algorithm does not produce theoretically valid knockoffs. We then turn to evaluating the gap in covariance estimation. In a Gaussian setting, to ensure exchangeability, it is necessary to have

\begin{align*}
    \left(X_1, \ldots, X_p, \tilde{X}_1, \ldots, \tilde{X}_p\right)\sim \mathcal{N}\left(
    \begin{pmatrix}
    \mu \\
    \mu
    \end{pmatrix}
    \begin{pmatrix}
    \Sigma & \Sigma - \mathrm{diag}(\mathbf{s}) \\
    \Sigma- \mathrm{diag}(\mathbf{s}) & \Sigma
    \end{pmatrix}\right),
\end{align*}

for some $\mathbf{s}$ such that the covariance matrix is still positive semidefinite (see \citealp{candes2018panning}). Using Gaussianity we can rewrite $X_j$ using $\mathbf{x}_{-j}$ and the residual from another i.i.d. sample $\mathbf{x}'$:

\begin{align*}
    \tilde{X}_j&:=\e{X_j\middle| X_{-j}}+X'_j-\e{X'_j\middle|X'_{-j}}\\
    &= \left(\mu_j+\Sigma_{j, -j}\Sigma_{-j,-j}^{-1}\left(X_{-j}-\mu_{-j}\right)\right)+X'_j-\left(\mu_j+\Sigma_{j, -j}\Sigma_{-j,-j}^{-1}\left(X'_{-j}-\mu_{-j}\right)\right)\\
    &=X'_j+\Sigma_{j,-j}\Sigma^{-1}_{-j,-j}\left(X_{-j}-X'_{-j}\right).
\end{align*}

From this formulation, we observe that the mean, variance, and covariance between knockoff and original covariate is preserved:
\begin{itemize}
    \item $
        \e{\tilde{X}_j}=\e{X'_j+\Sigma_{j,-j}\Sigma^{-1}_{-j,-j}\left(X_{-j}-X'_{-j}\right)}=\mu_j.
    $
    \item For any $j\in [1, p]$:
    \begin{align*}
        \e{\left(\tilde{X}_j-\mu_j\right)^2}&=\e{\left(X'_j-\mu_j+\Sigma_{j,-j}\Sigma^{-1}_{-j,-j}\left(X_{-j}-X'_{-j}\right)\right)^2}\\
        &=\e{\left(X'_j-\mu_j\right)^2}+\Sigma_{j,-j}\Sigma^{-1}_{-j,-j}\e{\left(X_{-j}-X'_{-j}\right)\left(X_{-j}-X'_{-j}\right)^\top}\Sigma^{-1}_{-j,-j}\Sigma_{-j,j}\\&\quad+2\e{\left(X'_j-\mu_j\right)\Sigma_{j,-j}\Sigma_{-j,-j}^{-1}\left(X_{-j}-X'_{-j}\right)}\\
        &=\Sigma_{j,j}+2\Sigma_{j,-j}\Sigma^{-1}_{-j,-j}\Sigma_{-j,-j}\Sigma^{-1}_{-j,-j}\Sigma_{-j,j} - 2\Sigma_{j,-j}\Sigma^{-1}_{-j,-j}\Sigma_{-j,j}
        \\
        &=\Sigma_{j,j}.
    \end{align*}
    \item Without loss of generality -- the same computation works for any $j\neq l\in \{1,\ldots,p\}$ -- we derive $\operatorname{Cov}(\tilde{X}_1, X_2)$: 
    \begin{align*}
        \e{\left(\tilde{X}_1-\mu_1\right)(X_2-\mu_2)}
        &=\e{\left(X'_1-\mu_1+\Sigma_{1,-1}\Sigma^{-1}_{-1,-1}\left(X_{-1}-X'_{-1}\right)\right)(X_2-\mu_2)}\\
        &=\Sigma_{1,-1}\Sigma^{-1}_{-1,-1}\e{\left(X_{-1}-X'_{-1}\right)(X_2-\mu_2)}\\
        &=\Sigma_{1,-1}\Sigma^{-1}_{-1,-1}\Sigma_{-1,2}\\
        &=\Sigma_{1,-1}\Sigma^{-1}_{-1,-1}\Sigma_{-1,-1}\left(\mathbf{1}, \mathbf{0}, \ldots, \mathbf{0}\right)\\
        &=\Sigma_{1,2},
    \end{align*}
    where in the second-to-last line we have used that $\Sigma_{-1,2}$ is the first column of $\Sigma_{-1,-1}$, and therefore we could rewrite it as 
    $\Sigma_{-1,-1}\left(\mathbf{1}, \mathbf{0}, \ldots, \mathbf{0}\right)$.
\end{itemize}

Without loss of generality, and for readability, we take the first and the second covariates of the knockoff and compute the covariance between them to show that it is not $\Sigma_{1,2}$ as it should be. Indeed, with this procedure, we could either add the residual from the same individual to each coordinate of the knockoff or add different and independent ones. We are going to show that in both cases, the covariance is not the desired one:
\paragraph{Residuals from independent samples:} Let $X'\overset{\mathrm{i.i.d.}}{\sim}X''$, then we have that the covariance between the first and second knockoff is given by:
\begin{align*}
    \mathbb{E}&\left[\left(\tilde{X}_1-\mu_1\right)\left(\tilde{X}_2-\mu_2\right)\right]\\&=\e{\left(X'_1-\mu_1+\Sigma_{1,-1}\Sigma_{-1,-1}^{-1}\left(X_{-1}-X'_{-1}\right)\left(X''_2-\mu_2+\Sigma_{2,-2}\Sigma_{-2,-2}^{-1}\left(X_{-2}-X''_{-2}\right)\right)\right)}\\
    &=\e{\Sigma_{1,-1}\Sigma_{-1,-1}^{-1}\left(X_{-1}-X'_{-1}\right)\Sigma_{2,-2}\Sigma_{-2,-2}^{-1}\left(X_{-2}-X''_{-2}\right)}\qquad\text{using $X'\indep X''$}\\
&=\Sigma_{1,-1}\Sigma_{-1,-1}^{-1}\e{\left(X^{-1}-X'_{-1}\right)\left(X_{-2}-X''_{-2}\right)^{\top}}\Sigma_{-2,-2}^{-1}\Sigma_{-2,2}\qquad\text{using $a^\top=a$ for $a\in \mathbb{R}$}\\
&=\Sigma_{1,-1}\Sigma_{-1,-1}^{-1}\Sigma_{-1,-2}\Sigma_{-2,-2}^{-1}\Sigma_{-2,2}.
\end{align*}

To see that this does not coincide with $\Sigma_{1,2}$, we may observe the simple two-dimensional unit variance random variable with covariance $\rho$. In this case, we have that $\Sigma_{1,-1}=\rho$, $\Sigma_{-1,-1}^{-1}=1$, $\Sigma_{-1,-2}=\rho$, $\Sigma^{-1}_{-2,-2}=1$ and $\Sigma_{-2,2}=\rho$, so $$\mathrm{Cov}\left(\tilde{X}^1, \tilde{X}^2\right)=\rho^3.$$

\paragraph{Residuals from the same sample:} When the knockoffs are constructed using the residuals from the same sample, the covariance is given by: 
\begin{align*}
    \mathbb{E}&\left[\left(\tilde{X}_1-\mu_1\right)\left(\tilde{X}_2-\mu_2\right)\right]\\
    &=\e{\left(X'_1-\mu_1+\Sigma_{1,-1}\Sigma_{-1,-1}^{-1}\left(X_{-1}-X'_{-1}\right)\left(X'_2-\mu_2+\Sigma_{2,-2}\Sigma_{-2,-2}^{-1}\left(X_{-2}-X'_{-2}\right)\right)\right)}\\
    &=\e{\left(X'_1-\mu_1\right)\left(X'_2-\mu_2\right)}+\e{\Sigma_{2,-2}\Sigma_{-2,-2}^{-1}\left(X_{-2}-X'_{-2}\right)\left(X'_1-\mu_1\right)}\\&\qquad+\Sigma_{1,-1}\Sigma_{-1,-1}^{-1}\e{\left(X_{-1}-X'_{-1}\right)\left(X'_2-\mu_2\right)}\\&\qquad+\e{\Sigma_{1,-1}\Sigma_{-1,-1}^{-1}\left(X_{-1}-X'_{-1}\right)\Sigma_{2,-2}\Sigma^{-1}_{-2,-2}\left(X_{-2}-X'_{-2}\right)}.
\end{align*}
The first term gives $\Sigma_{1,2}$. The second term can be simplified as
\begin{align*}
    \e{\Sigma_{2,-2}\Sigma_{-2,-2}^{-1}\left(X_{-2}-X'_{-2}\right)\left(X'_1-\mu_1\right)}&= -\Sigma_{2,-2}\Sigma_{-2,-2}^{-1} \Sigma_{-2,1}\\&=-\Sigma_{2,-2}\Sigma_{-2,-2}^{-1}\Sigma_{-2,-2}\left(\mathbf{1},\mathbf{0}, \ldots,\mathbf{0} \right)=-\Sigma_{2,1}.
\end{align*}

Similarly, the third term provides
\begin{align*}
    \Sigma_{1,-1}\Sigma_{-1,-1}^{-1}\e{\left(X_{-1}-X'_{-1}\right)\left(X'_2-\mu_2\right)}&=-\Sigma_{1,-1}\Sigma_{-1,-1}^{-1}\Sigma_{-1,2}\\
    &=-\Sigma_{1,-1}\Sigma_{-1,-1}^{-1}\Sigma_{-1,-1}\left(\mathbf{1}, \mathbf{0}, \ldots, \mathbf{0}\right)=-\Sigma_{1,2}.
\end{align*}

Finally, the last term can be simplified as
\begin{align*}
    \mathbb{E}&\left[\Sigma_{1,-1}\Sigma_{-1,-1}^{-1}\left(X_{-1}-X'_{-1}\right)\Sigma_{2,-2}\Sigma^{-1}_{-2,-2}\left(X_{-2}-X'_{-2}\right)\right]\\&=\Sigma_{1,-1}\Sigma_{-1,-1}^{-1}\e{\left(\mathbf{x}_{-1}-X'_{-1}\right)\left(X_{-2}-X'_{-2}\right)^\top}\Sigma_{-2,-2}^{-1}\Sigma_{-2,2}\\
    &=2\Sigma_{1,-1}\Sigma_{-1,-1}^{-1}\Sigma_{-1,-2}\Sigma_{-2,-2}^{-1}\Sigma_{-2,2}.
\end{align*}

Therefore, combining the previous, we obtain 
\begin{align*}
    \mathbb{E}&\left[\left(\tilde{X}_1-\mu_1\right)\left(\tilde{X}_2-\mu_2\right)\right]= -\Sigma_{1,2}+2\Sigma_{1,-1}\Sigma_{-1,-1}^{-1}\Sigma_{-1,-2}\Sigma_{-2,-2}^{-1}\Sigma_{-2,2}.
\end{align*}

Similarly as in the independent residuals setting, we can take the two dimension unit variance random variable to see that this will not coincide in general with the covariance $\rho$. Indeed, in this case, $$\mathrm{Cov}(\tilde{X}_1, \tilde{X}_2)=-\rho+2\rho^3.$$

In this example, for both independent and dependent residuals, although the correlation between the knockoff variables are close to the original one when the correlation is either small or large, they do not coincide; therefore, their parallel knockoffs are not theoretically valid.

\begin{figure}

\centering
\includegraphics[width=0.99\linewidth]{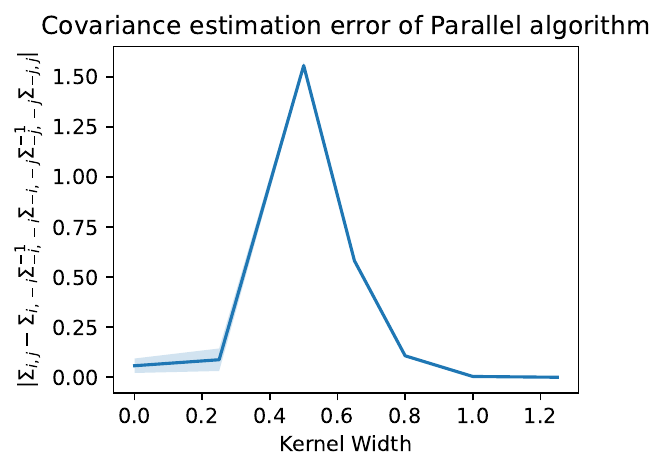}
\caption{\textbf{Covariance estimation error of Parallel generation.} We compute the difference between the correct covariance term $\Sigma_{1, 2}$ and the one produced by the parallel method, i.e. $\Sigma_{1,-1}\Sigma_{-1,-1}^{-1}\Sigma_{-1,-2}\Sigma_{-2,-2}^{-1}\Sigma_{-2,2}$. We vary the kernel width parameter to explore different correlation levels in the data. Notice that $\Sigma_{1, 2}$ is correctly estimated in regimes with practically no correlation and in regimes with very high correlation. The approximation is the least accurate for intermediate correlation levels.
}%
\label{fig:cov-error}
\end{figure}

In Figure \ref{fig:cov-error}, we use the setup of section \ref{sec:results} and explore a range of correlation levels by varying the kernel width parameter. The approximation error is practically zero when the data presents strong correlation or none at all. The error is more substantial in intermediate cases.

\end{appendix}

\end{document}